\documentclass[doublecol]{epl2} 

\title{Nonlinear Mach-Zehnder-Fano interferometer}
\shorttitle{Nonlinear Mach-Zehnder-Fano interferometer} 

\author{Yi Xu\inst{1,2} \and Andrey E. Miroshnichenko\inst{1}\thanks{E-mail: \email{\email{aem124@physics.anu.edu.au}}} }
\shortauthor{Yi Xu \etal}

\institute{                    
  \inst{1} Nonlinear Physics Centre and Centre for Ultra-high bandwidth Devices for Optical Systems (CUDOS),\\ Australian National University, Canberra ACT 0200, Australia\\
  \inst{2} Laboratory of Photonic Information Technology, School for Information and Optoelectronic Science and Engineering,\\ South China Normal University, Guangzhou 510006, P.R. China\\
}

\abstract{
We demonstrate that the interaction of loop and nonlinear Fano resonances results in a formation of hybrid resonant states in Mach-Zehnder type interferometers, providing with opportunities for an advanced phase manipulation. The nonlinear response of such structures can be greatly enhanced, leading to a low threshold 100$\%$ switching operation. We further propose one of the possible realizations  based on nonlinear photonic crystal circuits, suitable for optimal all-optical switching.}

\begin{document}

\maketitle

\section{Introduction}
Mach-Zehnder interferometer (MZI) is a key component in many branches of physics because of its ability to manipulate a coherent signal~\cite{MZ}. By coupling a resonator to the MZI can further increase the phase sensitivity of the coherent manipulation~\cite{boyd1,boyd2}. The enhanced all-optical switching~\cite{boyd1} and the bistability~\cite{ying} have been demonstrated in a coupled ring-resonator Mach-Zehnder interferometer, which provides the possibility for the effective and coherent control by using a nonlinear resonator. 

\begin{figure}[t]
\centerline{\includegraphics[width=8cm]{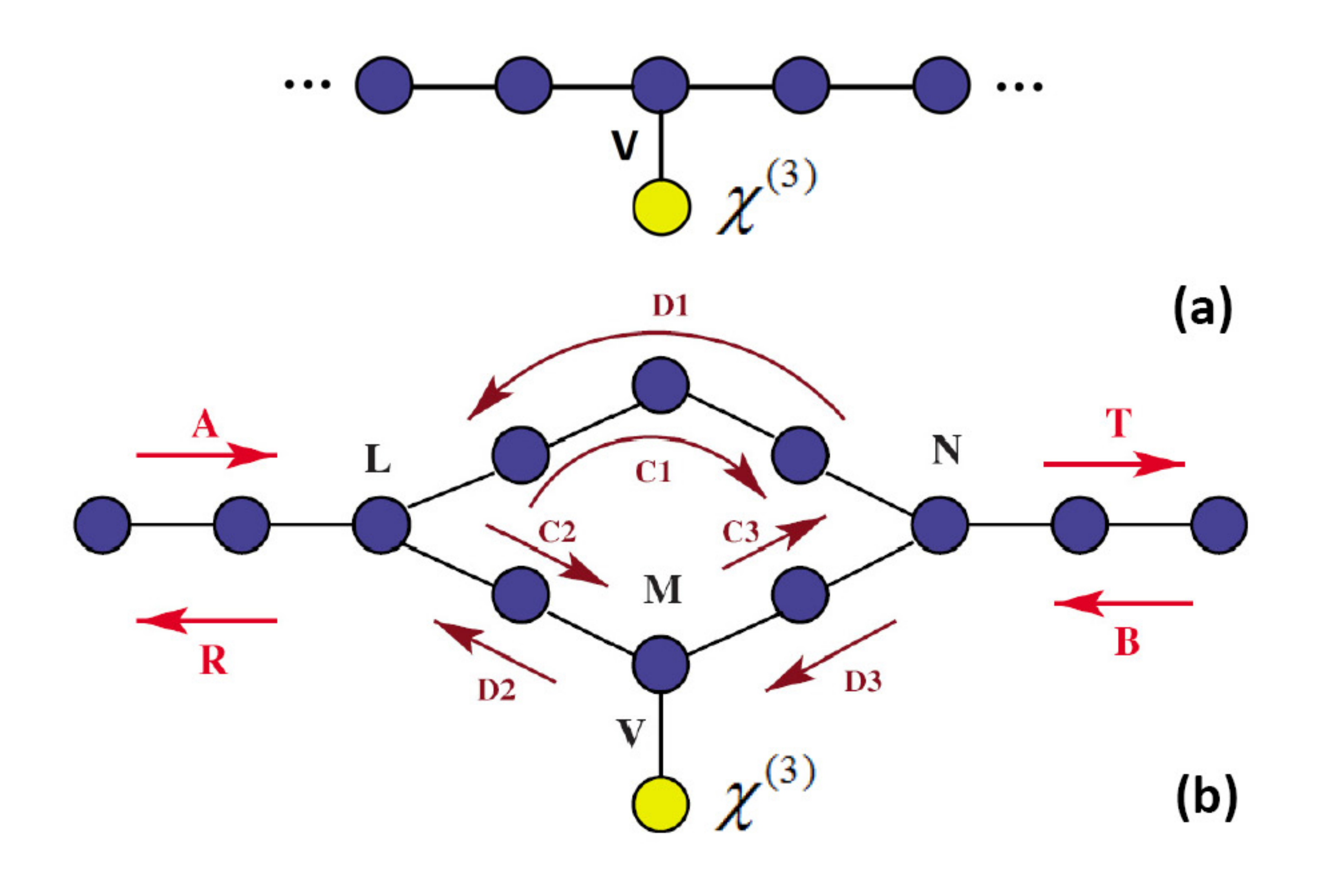}}
\caption{\label{fig:fig1} 
(color online)(a) and (b) are the generic discrete models for the system exhibiting Fano resonance. $ A $ and $ B $ are the incoming waves, $ T $ and $ R $ are scattered amplitudes. $ C_{i} $ and $ D_{i} $ are the forward and backward scattered waves. $ V $ is the coupling strength between the nonlinear Fano defect(located by $ M $) and the site $ A_{M} $ in the linear chain. $ L $ and $ N $ specify the number of sites. Without loss of generality, we choose L=1 which provides the phase reference while $ N $ represents the length of MZFI arm including two junction sites.
}
\end{figure}

Recently, we have introduced the concept of Mach-Zehnder-Fano interferometer (MZFI)~\cite{mzfi} providing with unique physical property that can not be found in a macroscopic resonator enhanced MZI~\cite{boyd1,boyd2,ying,asym}.
The MZFI allows us to manipulate the interaction of different types of resonances which leads to the formation of a novel hybrid Fano-like resonant states~\cite{yi_arxiv}. Furthermore, the counterpart of the ring-resonator-coupled Mach-Zehnder interferometer in the microscopic scale, i.e. MZFI based on a photonics crystal (PhC) platform seems to be more promising for future application owing to small volume compared with the macroscopic resonators. Recent advantages in PhCs fabrication technology~\cite{noda}, allow us to achieve ultra high-Q cavities facilitating low threshold nonlinear bistability~\cite{pre2002,yanik1,Martin,notomi1,yang}. Indeed, the manifested optical bistable state is the nonlinear Fano resonance~\cite{aem_rev}. As a result, the systems supporting Fano resonances~\cite{fano,aem_rev,B_review}, associated with an asymmetric scattering profile, attract a significant attention recently. It's their unique property that allows us to achieve an optimal high extinction ratio, large modulation depth and lowest threshold nonlinear switching~\cite{sfan,asym}. 

The aim of this Letter is to introduce and demonstrate unique properties of the nonlinear MZFI, which originate from the excitation of the nonlinear hybrid Fano resonances. Such resonant states appear because of the interaction between MZI loop's resonance and nonlinear Fano resonances of the side-coupled defect. It provides with an enhanced nonlinear response and optimal conditions for low threshold dynamic bistability. As a particular realisation of our model we provide with a PhC circuit example, which supports our results.

\begin{figure*}[t]
\centerline{\includegraphics[width=16.5cm]{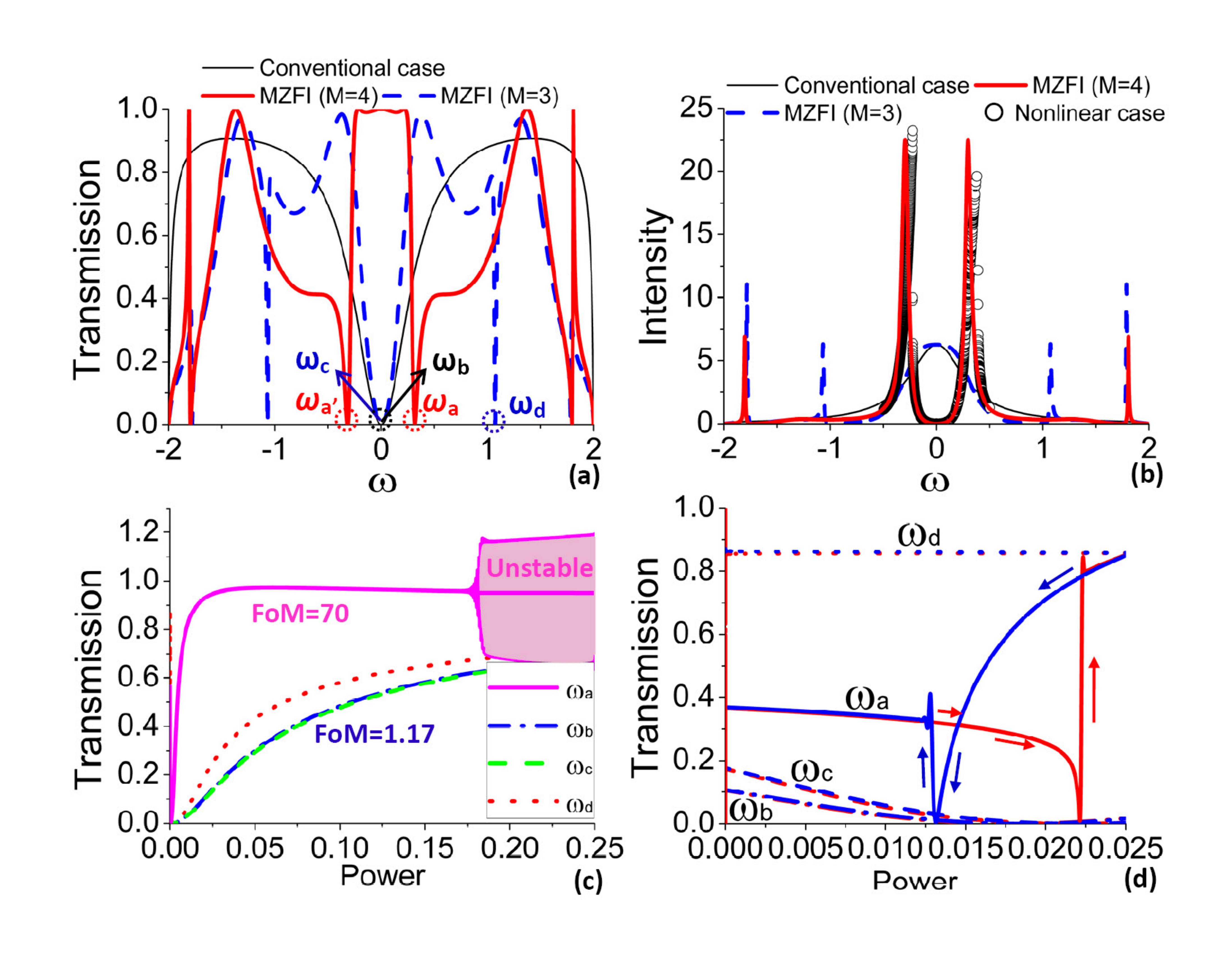}}
\caption{\label{fig:fig2} 
(color online)(a) and (b) are the transmission of the linear MZFI and the intensity of the defect, respectively. Here, $ L=1 $ and $ N=7 $. The circles in (b) represent the normalized intensity in the nonlinear Fano defect when the input power $ I=0.01 $ and $ \chi^{(3)}=1 $. (c) nonlinear response of different resonances marked $ \omega_{a-d} $ at Fig.~\ref{fig:fig2} (a). Solid pink line stands for the hybrid Fano resonance marked $ \omega_{a} $ in Fig.~\ref{fig:fig2} (a), dotted red line presents the resonance of the MZI's loop marked $ \omega_{d} $ while dashed green line and dashed-dotted blue line represents the resonance similar with the one in resonator enhanced MZI~\cite{boyd1,boyd2,ying,asym}(marked $ \omega_{c} $) and conventional one(marked $ \omega_{b} $), respectively. The shaded pink region represents the area of the dynamic modulation instability initialled by a pulse of $ t_{0}=1.3\times 10^{4} $, $ W=5\times 10^{3} $  and $ I_{0}=0.25 $. (d) Double bistability at the frequency $ \omega=0.43 $ with detuning $ \Delta\omega=0.11 $ respecting to $ \omega_{a} $. Here, $ t_{0}=1.3\times 10^{4} $, $ W=5\times 10^{3} $ and $ I_{0}=0.025 $. The arrows indicate the upward and downward nonlinear Fano bistable operation. The responses of $ \omega_{b-d} $ for the same detuning value are also presented.
}
\end{figure*}
\section{Discrete nonlinear MZFI model}
We start our analysis from a generic discrete model~\cite{aem_rev}, which can describe the dynamics of MZFI like structures. By using the modified Fano-Anderson model, dynamical equations describing the nonlinear MZFI system shown in Fig.~\ref{fig:fig1} (b) can be written as follows:
\begin{equation}
\label{eq.1}
\begin{array}{rcl}
i\dot{\psi}_n=\sum_{k}\psi_{k}+\delta_{n,M}V\varphi_d\;, \\
i\dot{\varphi}_d=E_{d}\varphi_d+\chi^{(3)}|\varphi_d|^2\varphi_d+V\psi_M\;,
\end{array}
\end{equation}
where $ \psi_{n} $ represents the linear chain with complex field amplitude, $ \varphi_{d} $ stands for the Fano defect, $ M $ gives the location of the Fano defect in the arm, $ k $ is the total number of the nearest neighbour sites in the chain($ k=3 $ in the Y-junction while $ k=2 $ at the others) and we neglect the nonlocal interactions in this paper, $ \chi^{(3)} $ is the cubic nonlinear parameter, $ V $ is the coupling strength between the chain and the Fano defect, $ E_{d} $ is the eigenfrequency of the Fano defect. 

As it was demonstrated in Ref.~\cite{aem1}, the side-coupled Fano defect acts as an effective scattering potential, whose strength depends on the input frequency $\omega$. In the presence of nonlinearity the effective scattering potential of the Fano defect becomes input intensity dependent, which brings extra flexibility to tune the Fano resonances of the MZFI. 

To find static solutions of system (1) we employed S-matrix approach. The S-matrices of the Y-junction can be obtained by applying scattering boundary condition at three branches respectively~\cite{aem2}. A set of nonlinear equations which describes the stationary nonlinear response of the system is as follow:

\begin{eqnarray}
\label{eq.2}
&\left(
  \begin{array}{c}
  R\\C_1\\C_2
  \end{array}
\right)&
  ={\mathbf S}^Y_L
  \left(
  \begin{array}{c}
  A\\D_1\\D_2
  \end{array}
\right)\;,\;
\left(
  \begin{array}{c}
  T\\D_1\\D_3
  \end{array}
\right)
  ={\mathbf S}^Y_{-N}
  \left(
  \begin{array}{c}
  B\\C_1\\C_3
  \end{array}
\right),
\nonumber\\
&\left(
  \begin{array}{c}
  D_2\\C_3
  \end{array}
\right)&
  ={\mathbf S}_M
  \left(
  \begin{array}{c}
  C_2\\D_3
  \end{array}
\right),
\nonumber\\
&X_{d}&=\frac{A_{M}V}{(\omega-E_{d}-\lambda|X_d|^2)}\nonumber\\
\end{eqnarray}

where 

$ {\mathbf S}^Y_K=\left(
  \begin{array}{ccc}
  e^{-2iKq}r_Y&t_Y&t_Y\\
  t_Y&e^{2iKq}r_Y&e^{2iKq}t_Y\\
  t_Y&e^{2iKq}t_Y&e^{2iKq}r_Y
  \end{array}
\right),
\nonumber\\
{\mathbf S}_M=
\frac{1}{\varepsilon -2i \sin q}
\left(
  \begin{array}{cl}
  -\varepsilon e^{-2iMq}&-2i \sin q\\-2i \sin q&-\varepsilon e^{2iMq}
  \end{array}
\right),\nonumber\\
\varepsilon=\frac{V^2}{(\omega-E_{d}-\lambda|X_d|^2)} $

As can be seen from the complex defect field $ X_{d} $, the nonlinearity has an effect on the phase and amplitude of the Fano defect and the corresponding strengths in turn depend on the power of the Fano defect. At the same time, the nonliear response of the Fano defect would give a feedback to the scattering waves in the arms of MZFI (read as the \textit{nonlinear scattering potential} $ \varepsilon $) of which forms a complex nonlinear resonance system. Then, such nonlinear MZFI, whose nonlinear response can not be mapped from the standard nonlinear Fano resonant system~\cite{aem1}, is sophisticated even in the linear region~\cite{yi_arxiv}. 

We study two cases, where the Fano defect is placed symmetrically and asymmetrically. The detail parameters can be found in the caption of Fig.~\ref{fig:fig2}. These two cases allow us to excite different sets of MZI loop's modes, which have nonzero overlap with the coupled site~\cite{yi_arxiv}. The scattering of the conventional Fano resonance geometry [see Fig.~\ref{fig:fig1}(a)] is shown by dotted line in Fig.~\ref{fig:fig2}(a) as a reference. In the symmetric case, the transmission in the centre of the propagation band resembles a step-function. In such a case, the eigenfrequency of the Fano defect is in the vicinity of MZI loop's resonance, where a hybrid Fano resonance is formed. At the same time, such resonance has the highest intensity at the defect site among other resonances because of the interference of two Fano-like resonances~\cite{yi_arxiv}. Intuitively, the high contrast step function like transmission together with high resonant intensity of the nonlinear defect are positive factors for further enhancment of the nonlinear behaviour~\cite{box}. We, thus, investigate the nonlinear response of the system (1) at these specific resonances. 
 
The stationary nonlinear switching of four resonances marked $ \omega_{a-d} $(corresponding to transmission minima) in Fig.~\ref{fig:fig2} (a) are shown in Fig.~\ref{fig:fig2} (c). It should be pointed out that the maximum of the Fano defect's intensity is in between the transmission dip and tip because of the sharp asymmetric line shapes~\cite{aem_rev}. We emphasize that the resonance $ \omega_{d} $ is excited due to a symmetry breaking by side-coupled Fano defect~\cite{yi_arxiv} and it is \textit{not} the eigenfrequency of the defect. At the same time, the nonlinear response at $ \omega_{a} $ is greatly reduced by the step-function like linear transmission compared to resonance $ \omega_{b}$ and $ \omega_{c}$. If we define a figure of merit as $ FoM=\Delta T/P_{th} $, where $ \Delta T=\max\limits_{P_{in}}[T(P_{in})-T(0)] $ refers to the maximum contrast of the transmission and $ P_{th} $ represents the necessary input power to pull the transmission of the system up to $ 90 \% $ of the $\Delta T $. The FoM of resonance $ \omega_{a}$, which describes both the enhanced transmission contrast in the linear case and the reduction of the switching power, can be enhanced more than 60 times compared to a given defect supporting conventional Fano resonance $ \omega_{b}$~\cite{aem1}. The nonlinear responses at $ \omega_{b}$ and $ \omega_{c} $ are similar because they are originated from the eigenfrequency of the Fano defect.  

To study the dynamical switching we employed high accuracy Crank-Nicolson method~\cite{CK} with suitable absorption boundary condition~\cite{dtbc}. The dynamic nonlinear response of the system at specified resonances is obtained by a long pulse with $ I=I_{0}\exp(-(t-t_{0})^{2}/W^{2})\sin(\omega t) $, where $ W $ is the pulse width, $ \omega $ is set at the same frequencies as $ \omega_{a-d} $. Compared to the stationary one, the dynamic solution predicts similar nonlinear response except for the case of $ \omega_{a} $. One can see that transmission cannot be well defined at frequency $ \omega_{a} $ above a certain threshold of the input power, indicated by the shaded region. This is caused by the modulational instability of the Fano resonances~\cite{dynamic}, which also indicates an enhanced nonlinear response. It is similar to the modulational instability of waves scattered by a nonlinear centre~\cite{mi}. Usually, it happens in the finite interval of frequencies . The stability analysis of our system with the nonlinear hybridization between Fano resonances  gives similar results to the conventional nonlinear Fano resonance with the same instability regions~\cite{dynamic}. Note here, that dynamical instability occurs independently from bistability conditions.

\begin{figure}[htb]
\centerline{\includegraphics[width=8cm]{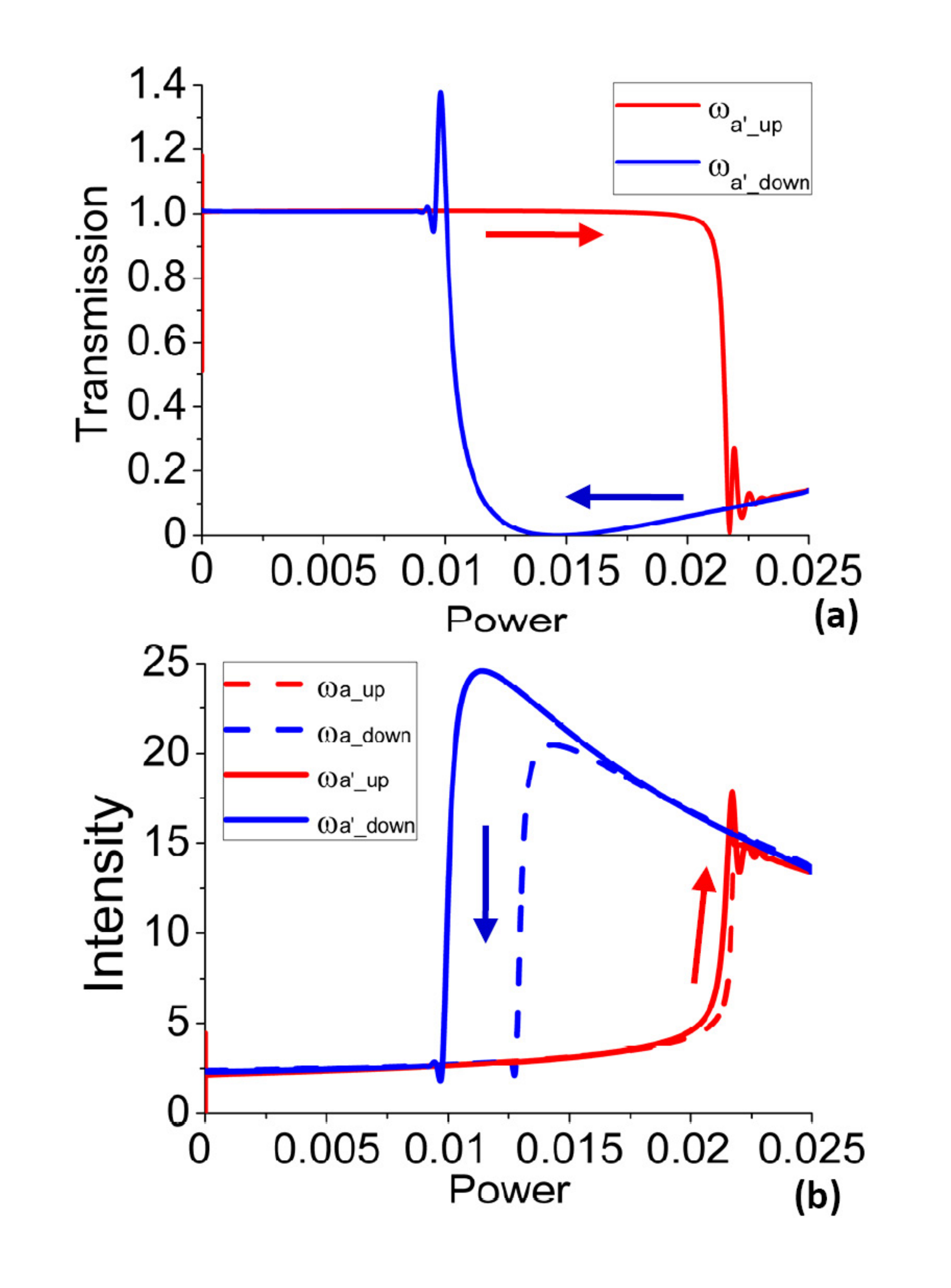}}
\caption{\label{fig:fig3} 
(color online) (a) Typical perfect bistability at the frequency $ \omega=-0.2 $ with respect to $ \omega_{a'} $. Here, $ t_{0}=1.3\times 10^{4} $, $ W=5\times 10^{3} $ and $ I_{0}=0.025 $. (b) Intensity of the Fano defect with detuning $ \Delta \omega =0.11 $ with respect to $ \omega_{a,a'} $.
}
\end{figure}
Figure~\ref{fig:fig2} (d) demonstrates a butterfly like double bistability obtained by dynamic pulse feeding, where the operation frequency is referred to $ \omega_{a}=0.43 $.  Such a hysteresis loop at $ \omega_{a} $ shows interesting properties that one loop is clockwise while the other is anti-clockwise. 
The non-uniform shift of the normalized intensity in the nonlinear Fano defect [see balls in Fig.~\ref{fig:fig2} (b)] leads to the double bistable operation. As can be seen from this figure, that such a detuning is not enough to initiate a bistable response for other case of $ \omega_{b} $, $ \omega_{c} $, and $ \omega_{d} $. Thus, the nonlinear hybridization of various resonant states offers a unique opportunity to manipulate the nonlinear Fano resonances. According to the stationary normalized intensity of the defect [see Fig.~\ref{fig:fig2} (b)], we can obtain further nonlinear enhancement by working with $ \omega_{a'} $. Figure~\ref{fig:fig3} (a) shows an enhanced bistability when the operation frequency is set at $ \omega=-0.21 $. This kind of perfect bistable state benefits from the sharp step function like linear transmission and the further enhanced intensity of the nonlinear defect. The intensities of the Fano defect [see Fig.~\ref{fig:fig3}(b)] which demonstrate the further enhancement of the intensity in the Fano defect are in accord with the static result [see Fig.~\ref{fig:fig2}(b)]. The oscillation between two bistable state is the process that one bistable state transfers to the other one and it is the properties of the dynamical bistability~\cite{dynamic}. 
\begin{figure}[htb]
\centerline{\includegraphics[width=7.5cm]{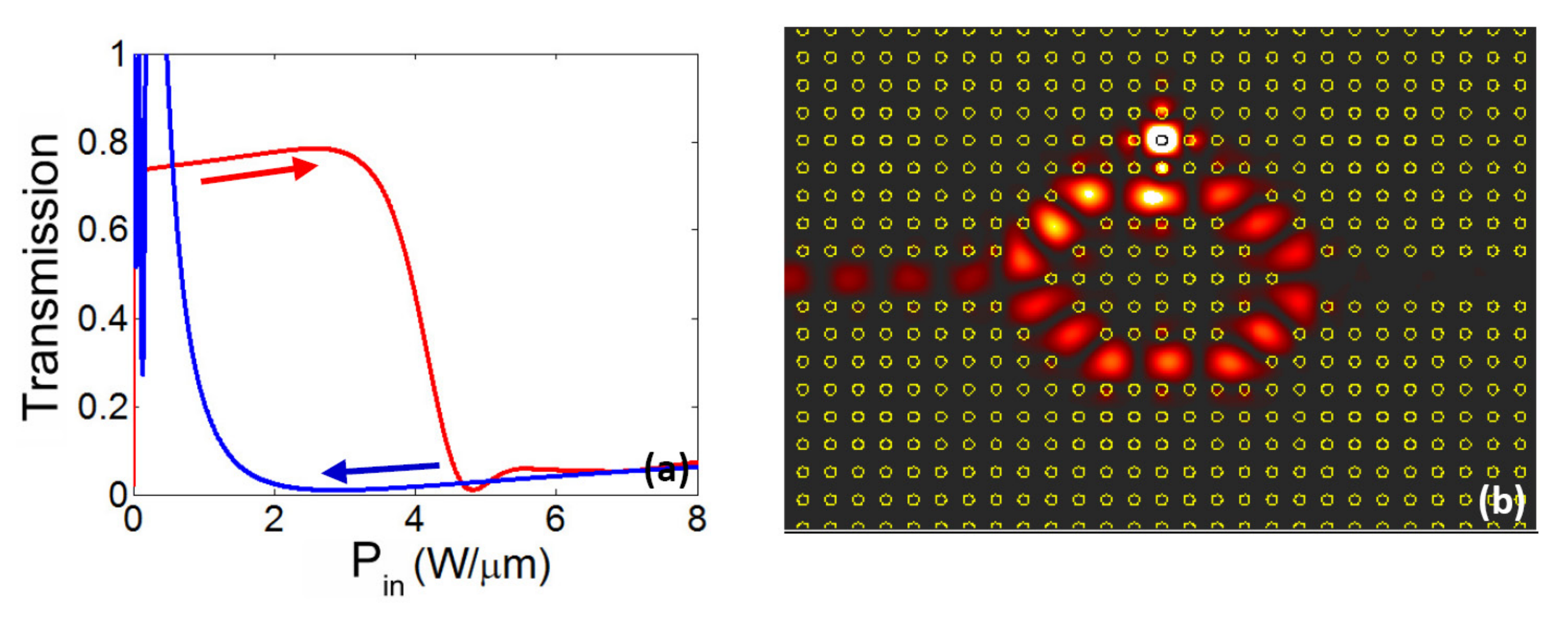}}
\caption{\label{fig:fig4} 
(color online)(a) Bistable operation obtain by pulse excitation in nonlinear FDTD experiment; (b) electric field distribution ($ \vert E\vert^{2} $) at the upward dynamic switching point. 
}
\end{figure}

\section{Photonic Crystal based nonlinear MZF interferometer}
Using the modified Fano-Anderson model, we have proved that the nonlinear MZFI can be considered as suitable candidate to realize the enhanced nonlinear response and manipulate the dynamical bistability. We thus suggest a PhCs platform as one of the possibilities to realize our idea, while the results above can be applied to other varieties of nonlinear discrete system. The PhC structure is shown in Fig.~\ref{fig:fig4} (b) and It is formed by dielectric rods embedded in air(the radius of dielectric rods is $r=0.19a $ where $ a $ is the lattice constant and refractive index $ n = 3.14$) except for the nonlinear Fano defect (polymer rod $ n=1.59 $ and the third-order nonlinearity susceptibility of the polymer is $ \chi^{(3)}=1.14\times 10^{-12} cm^{2}/W $). We use the nonlinear FDTD method to solve the Maxwell's equations. The bistable response shown in Fig.~\ref{fig:fig4} (a) is obtained by the response of a Gaussian pulse with input frequency $ f=0.373$  $2\pi c/a$ and duration $ 30 $ picoseconds. The profile is similar to the theoretical result in Fig.~\ref{fig:fig3}. Figure~\ref{fig:fig4} (b) shows the transient electric field distribution ($ \vert E\vert^{2} $) exactly at the downward switching point (from on to off-state). Nearly perfect blocking of the input pulse demonstrates the dynamical shutting down operation by a pulse and successful suppression of the modulation instability. Both high intensities of the Fano defect and the loop are the signature of nonlinear switching involving unique hybrid Fano resonances in the MZFI~\cite{yi_arxiv}. The transmission contrast can be further enhanced by careful engineering the interaction between the eigen-Fano resonance and the MZI loop's resonance which takes place exactly at the propagation band center. 

\section{Conclusions}
In conclusion, we have investigated the interaction of loop and side-coupled Fano resonances in  nonlinear MZFI type structures. By introducing nonlinear Fano defects we are able to excite "dark" states of pure MZI, which results in the formation of very narrow hybrid resonant states. Both stationary and dynamic studies convince the superiority of enhanced nonlinear response and dynamical bistability in nonlinear MZFI. Our direct numerical simulations of two-dimensional PhCs confirm the theoretical predictions of the dynamic characteristics. We anticipate that such structures can be realized on different platforms for ultra-high sensitive sensing operations. The idea of nonlinear Fano resonances hybridization can also be generalized to the other nanoscale structure. Particularly, in the new emerging field of plasmonics and metamaterial where the electromagnetic wave is confined in subwavelength scale, such nonlinear resonance interaction could be a basic physical idea for designing the logic devices.

\acknowledgments
The authors thank Prof. Yuri Kivshar and Dr. Anton Desyatnikov for useful discussions. Y. Xu acknowledges the support from the China Scholarship Council and the Nonlinear Research Centre at ANU for their hospitality. The work of A. E. Miroshnichenko was supported by the Australian Research Council through Future Fellowship program.


\begin{thebibliography}{0}

\bibitem{MZ}
  \Name{Ernst M.}
  \Book{The Principles of Physical Optics}
  \Publ{Courier Dover Publications, Dover}
  \Year{2003}
  
\bibitem{boyd1}
  \Name{Heebner J. E. \and Boyd R. W.}
  \REVIEW{Opt. Lett.}{24}{1999}{847-849}.
  
\bibitem{boyd2}
  \Name{Heebner J. E. \it et. al.}
  \REVIEW{Opt. Lett.}{29}{2004}{769-771}.
  
\bibitem{ying}
  \Name{Lu Y. \it et. al.}
  \REVIEW{Opt. Lett.}{30}{2005}{3069-3071}.

\bibitem{mzfi}
  \Name{Miroshnichenko A. E. \and Kivshar Y. S.}
  \REVIEW{Appl. Phys. Lett.}{95}{2009}{121109}.
	
\bibitem{asym}
  \Name{Mario L. Y., Darmawan S. \and Chin M. K.}
  \REVIEW{Opt. Express}{14}{2006}{12770-12781}.

\bibitem{yi_arxiv}
  \Name{Xu Y. \and Miroshnichenko A. E.}
  \REVIEW{Phys. Rev. A}{84}{2011}{033828}.

\bibitem{noda}
  \Name{Akahane Y. \it et. al.}
  \REVIEW{Nature}{425}{2003}{944}.

\bibitem{pre2002}
  \Name{Solja$ \check{c} $i$ \acute{c} $ M. \it et. al.}
  \REVIEW{Phys. Rev. E}{66}{2002}{055601}. 

\bibitem{yanik1}
  \Name{Yanik M. F., Fan S. \and Soljacic M.}
  \REVIEW{Appl. Phys. Lett.}{14}{2003}{2739-2741}.

\bibitem{Martin}
  \Name{Solja$ \check{c} $i$ \acute{c} $ M. \it et. al.}
  \REVIEW{Opt. Lett.}{28}{2003}{637-639}. 

\bibitem{notomi1}
  \Name{Tanabe T. \it et. al.}
  \REVIEW{Opt. Lett.}{30}{2005}{2575-2577}. 

\bibitem{yang}
  \Name{Yang X. \it et. al.}
  \REVIEW{Appl. Phys. Lett.}{91}{2007}{051113}. 

\bibitem{fano}
  \Name{Fano U.}
  \REVIEW{Phys. Rev.}{124}{1961}{1866}. 
	
\bibitem{aem_rev}
  \Name{Miroshnichenko A. E., Flach S. \and Kivshar Y. S.}
  \REVIEW{Rev. Mod. Phys.}{82}{2010}{2257}.
  
\bibitem{B_review}
  \Name{Luk'yanchuk B., Zheludev N. I., Maier S. A., Halas N. J., Nordlander P., Giessen H., \and Chong C. T.}
  \REVIEW{Nat. Mater.}{9}{2010}{707-715}.

\bibitem{sfan}
  \Name{Fan S.}
  \REVIEW{Appl. Phys. Lett.}{80}{2002}{908-910}. 
	
\bibitem{aem1}
  \Name{Miroshnichenko A. E. \it et. al.}
  \REVIEW{Phys. Rev. E}{71}{2005}{036626}. 

\bibitem{aem2}
  \Name{Miroshnichenko A. E., Molina M. I. and Kivshar Y. S.}
  \REVIEW{Phys. Rev. E}{75}{2007}{046602}. 

\bibitem{box}
  \Name{Darmawan S., Landobasa Y. M. \and Chin M. K.}
  \REVIEW{Opt. Express}{15}{2007}{437-448}.

\bibitem{dynamic}
  \Name{Miroshnichenko A. E. \it et. al.}
  \REVIEW{Phys. Rev. A}{79}{2009}{013809}. 

\bibitem{CK}
  \Name{Crank J. \and Nicolson P.}
  \REVIEW{Proc. Camb. Phil. Soc.}{43}{1947}{50-67}.

\bibitem{dtbc}
  \Name{Arnold A., Ehrhardt M. \and Sofronov I.}
  \REVIEW{Commun. Math. Sci.}{1}{2003}{501}.

\bibitem{mi}
  \Name{Malomed B. A. and Azbel M. Y.}
  \REVIEW{Phys. Rev. B}{47}{1993}{10406}. 

\end{thebibliography}
\end{document}